\documentclass[prl %
 aps,
 amsmath,amssymb,
 reprint,%
author-numerical,%
]{revtex4-1}
\usepackage{booktabs}
\usepackage{lipsum}
\usepackage{multirow}
\usepackage{subfig}      

\usepackage{graphicx}
\usepackage{dcolumn}
\usepackage{bm}
\usepackage{float}
\usepackage{xcolor}
\usepackage{amsmath}
\usepackage[T2A]{fontenc}
\usepackage[utf8]{inputenc}
\usepackage{soul}
\usepackage{natbib}
\usepackage[colorlinks=true, allcolors=blue]{hyperref}

\begin{document}

\title{Measuring coherent dynamics of a superconducting qubit in an open waveguide}

    \author{A.~Sultanov}
	\affiliation{Leibniz Institute of Photonic Technology, D-07745 Jena, Germany}

    \author{E.~Mutsenik }
    \email{evgeniya.mutsenik@leibniz-ipht.de}
    \affiliation{Leibniz Institute of Photonic Technology, D-07745 Jena, Germany}

    \author{M.~Schmelz}
	\affiliation{Leibniz Institute of Photonic Technology, D-07745 Jena, Germany}

    \author{L.~Kaczmarek}
	\affiliation{Leibniz Institute of Photonic Technology, D-07745 Jena, Germany}

    		\author{G.~Oelsner}
	\affiliation{Leibniz Institute of Photonic Technology, D-07745 Jena, Germany}

	\author{U.~H\"ubner}
	\affiliation{Leibniz Institute of Photonic Technology, D-07745 Jena, Germany}
 
	\author{R.~Stolz}
	\affiliation{Leibniz Institute of Photonic Technology, D-07745 Jena, Germany}

        \author{E.~Il'ichev}
	\affiliation{Leibniz Institute of Photonic Technology, D-07745 Jena, Germany}

\begin{abstract}
We measured the relaxation and decoherence rates of a superconducting transmon qubit in a resonator-free setting. In our experiments, the qubit is coupled to an open coplanar waveguide such that the transmission of microwaves through this line depends on the qubit's state. To determine the occupation of the first excited qubit energy level, we introduced a two-pulse technique. The first applied pulse, at a frequency close to the eigenfrequency of the qubit, serves to excite the qubit. A second pulse is then used for probing the transition between the first and second excited energy levels. Utilizing this measurement technique allowed for the reconstruction of the relaxation dynamics and Rabi oscillations. Furthermore, we demonstrate the consistency between the extracted parameters and the corresponding estimations from frequency-domain measurements. 
 
\end{abstract}

\maketitle
The field of quantum communication and quantum information processing attracts considerable interest of experts to realize basic elements of quantum internet and processors. Superconducting quantum circuits \cite{Clarke2008, Krantz2019} have emerged as one of the most promising hardware platforms for these applications, owing to their scalability and strong coupling to external fields, which enable relatively fast operation.

In practice, circuit quantum electrodynamics (cQED) is widely used to implement these devices. The basic cell of cQED is a nonlinear superconducting circuit acting as a qubit, coupled to a waveguide resonator \cite{Blais2021}. Simple quantum algorithms have been demonstrated within this architecture, see e.g. \cite{DiCarlo2009, Kandala2017, Arute2020}. Moreover, the performance of processors with several dozen qubits was investigated. In particular, the experimental realization of quantum supremacy was announced \cite{Arute2019,Gao2025}. 

Recent developments in quantum technologies have led to new experimental implementations of the building blocks for the quantum-optics toolbox for microwave photons. It is based on the strong coupling achievable between microwave photons in a waveguide and qubits \cite{Astafiev2010}. For example, such resonator-free settings have been used for the generation \cite{Peng2016, Forn-Diaz2017} and detection \cite{Kono2018, Besse2018} of single microwave photons. The ability to distribute quantum entanglement by generating spatially entangled itinerant photons, an essential feature of future quantum networks, has been demonstrated \cite{Oliver2020}. To manipulate propagating photons a quantum router basically consisting of a set of coplanar waveguides coupled by superconducting qubits has been proposed \cite{Sultanov2020}. 

The efficiency of photon manipulation in a waveguide-qubit setup depends on the qubit decoherence rate. Standard measurement procedures based on dispersive qubit readout, which are commonly used in cQED, cannot be applied in this context due to the lack of a resonator. In this Letter, we report on a direct method to measure relaxation and dephasing qubit rates in such resonator-free setups.

The main idea of the experiment is to utilize transitions between the ground and first excited states with frequency $\omega_{01}$ as well as between the first and second excited states with frequency $\omega_{12}$. Basically, this is similar to the demonstration of electromagnetic induced transparency in cQED \cite{Abdumalikov2010}. Therein, driving the transition to the second excited state was used for manipulation of the microwave absorption at the frequency $\omega_{01}$ of a flux qubit. In that work, $\omega_{12}$ was not directly accessible for readout because of the large anharmonicity of the used flux qubit, thus $\omega_{12}$ laid outside of the measurement bandwidth. The low anharmonicity of transmon qubits, on the other hand, results in closely spaced transition frequencies, $\omega_{01}-\omega_{12} \ll \omega_{01}$. Therefore, both of these transitions can be easily probed with standard measurement setups by applying appropriate resonant pulses to the transmission line. This enables novel experimental protocols to be implemented using separate drive and readout pulses that initialize and verify the occupation of the first excited energy level of a qubit.

We have investigated a single flux-dependent transmon qubit \cite{Koch2007} capacitively coupled to a coplanar open waveguide with an on-chip DC-bias line. Manhattan-type technology was used to fabricate the qubits' Josephson junctions as well as capacitors and coplanar waveguides \cite{Schmelz2024}. To mitigate unwanted modes, aluminum air bridges were implemented on the chip. They were fabricated by a combination of contact lithography, e-beam evaporation, and wet etching, as described in detail in \cite{Kaczmarek2025}. The sample is mounted on the base of the dilution refrigerator at an ambient temperature of 10 mK. A description of the measurement setup is given in \cite{Supplementary}. 
The input signal is attenuated by -110~dB as a combination of -70~dB attenuation in the dilution refrigerator and an additional -40~dB at room temperature. This setting of attenuation allows us to cover a range of 1870 to 0.6 average photons per coupling period of the qubit device.

\begin{figure}[H]
\centering
\includegraphics[width=0.9\columnwidth]{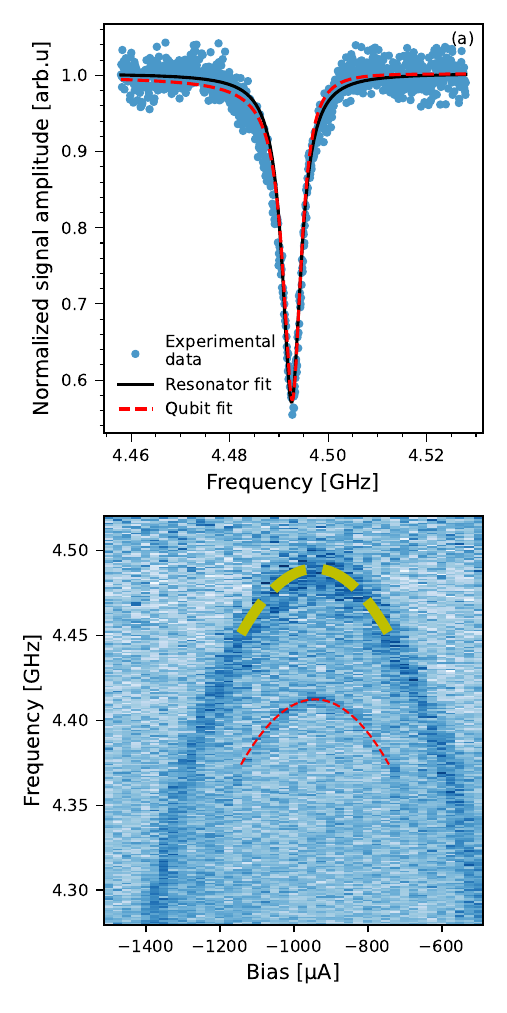}
 \caption{\label{fig:Qubit_spectra} Qubit's steady state responses. a) response at the sweet spot measured at $\Omega_p=2\pi\cdot 1.06$~MHz. Black solid and red dashed lines are the respective fits with the resonator \eqref{eq:circle} and qubit model fits \eqref{eq:qubit} b) qubit spectroscopy measured with $\Omega_p=2\pi \cdot 18.95 $~MHz, showing both $\omega_{01} = 2\pi f_{01}$ and $\omega_{02}/2 = 2\pi \frac{f_{02}}{2}$ transitions fitted by second order polynomials as described in the main text.}
 \end{figure}

To characterize the steady-state properties of the qubit at the sweet spot bias, we use a relatively long pulse duration (2~\textmu s) with an amplitude relating to the input power -145~dBm corresponding to an average photon number $N_{ph}=\frac{P_{in}\cdot 2 \mu s}{\hbar \omega_{01}}\approx2.1$ at the sample's input. Due to destructive interference, a dip at $\omega_{01}$ which is associated with the transition of the qubit to the first excited state, is observed in the transmission coefficient, see Fig.~\ref{fig:Qubit_spectra}a \cite{Astafiev2010}.
This response allows estimating the qubit parameters via fitting with two models:
\paragraph{resonator fit}
or circle fit \cite{Probst2015}, where the transmission through the qubit in the waveguide is treated as a notch-type resonator: 
\begin{equation}\label{eq:circle}
    S_{21}^{res}(\omega)=E(\omega) \left(1- \frac{\left(Q_l/Q_c\right) e^{i\phi}}{1+2iQ_l\left(\omega/\omega_{01}-1\right)}\right).
\end{equation}
Herein $E(\omega)$ are the characteristics of the environment/setup, $\frac{1}{Q_l}=\frac{1}{Q_c}+\frac{1}{Q_i}$, $Q_l$ is the loaded quality factor, $Q_c$ is the coupling quality factor, $Q_i$ is the internal quality factor of the resonator, $\phi$ is the impedance mismatch distortion and $i$ is the imaginary unit. This model gives a simple representation of the scattering process, where a distinction is made between radiative losses, defined by $Q_c$, and non-radiative losses, defined by $Q_i$. 
\paragraph{qubit fit}
or an equation, originally developed for a qubit in an open waveguide configuration \cite{Astafiev2010} and later updated \cite{Brehm2021, Hoi2011}, that directly relates the transmission coefficient to the qubit parameters: 
\begin{equation}\label{eq:qubit}
     S_{21}^{qb}(\omega)= 1- \frac{\Gamma_{10}}{2\gamma_{10}}\frac{1-i\frac{\omega-\omega_{01}}{\gamma_{10}}}{1+\left(\frac{\omega-\omega_{01}}{\gamma_{10}}\right)^2+\frac{\Omega_p^2}{\left(\Gamma_{10}+\Gamma_l\right)\gamma_{10}}},
\end{equation}
where $\Gamma_{10}$ is the radiative loss rate, defined by the coupling between the qubit and the waveguide, $\gamma_{10}=\frac{\Gamma_{10}}{2}+\frac{\Gamma_l}{2}+\Gamma_{\varphi}$ is the decoherence rate, where $\Gamma_{\varphi}$ is the pure dephasing rate, $\Gamma_l$ is the non-radiative relaxation rate and $\Omega_p$ is the Rabi driving rate.

A detailed comparison of the two fitting approaches establishes a direct link between qubit parameters and energy loss processes. In the weak-drive approximation, ${\Omega_p^2}\ll{\left(\Gamma_{10}+\Gamma_l\right)\gamma_{10}}$, Eq.~\ref{eq:qubit}  simplifies to the following form:
\begin{equation*}
     S_{21}^{qb}(\omega)= 1- \frac{\frac{\omega_{01}}{2\gamma_{10}}\frac{\Gamma_{10}}{\omega_{01}}}{1+i\frac{\omega_{01}}{\gamma{01}}\left(\omega/\omega_{01}-1\right)}.
\end{equation*}
From this expression it follows that the coupling quality factor is given as $Q_c=\frac{\omega_{01}}{\Gamma_{10}}$ and the internal quality factor is $Q_i=\frac{\omega_{01}}{\Gamma_l+2\Gamma_{\varphi}}$. This formulation allows for interpreting the qubit response analogously to a simple resonator, where the dissipation mechanisms are expressed in a form that facilitates understanding of the qubit's loss processes. For example, the energy transfer between a qubit and a waveguide is naturally dependent on the impedance mismatch. Thus, the coupling between the qubit and the waveguide $\Gamma_{10}$ is treated as a complex quantity during the fitting procedure that incorporates a phase factor $e^{i\phi}$. Similarly, irreversible energy losses, quantified by $Q_i$, clearly arise from both non-radiative relaxation and pure dephasing processes.

The measured data as well as the fits based on Eq.~\ref{eq:circle} and Eq.~\ref{eq:qubit} are presented in Fig.~\ref{fig:Qubit_spectra}. The experimental data are therein normalized to the amplitude of the transmission coefficient measured at frequencies where the influence of qubit dynamics is negligible. Both fits are consistent with the experimental data. The reconstructed parameters are listed in Table~\ref{tab:Compare_table}. The agreement between the two fits indicates that the used drive power is efficiently small compared to the loss rates. This observation is further supported by the Rabi drive calibration shown below.

The implementation of the time-domain measurements described above requires knowledge of $\omega_{12}$. To measure it, we performed qubit spectroscopy in the presence of an external flux bias with a strong drive as -120~dBm or $N_{ph}\approx700$ per qubit cycle. The latter is required to induce two-photon transitions at $\omega_{02}/2$. We subtracted the average spectra to enhance the contrast of the plot since it degrades under the strong drive, as it follows from Eq.~\ref{eq:qubit}. This result is shown in Fig.~\ref{fig:Qubit_spectra}b.
By making use of a conventional second-order polynomial fit, we get two polynomials with these coefficients $-960\left[\frac{Hz}{{\mu A}^2}\right]\cdot Bias^2 \left[{\mu A}^2\right]+4.49(4.4135)\left[GHz\right]$. Thus at the sweet spot the transition frequencies are $\omega_{01} =2\pi\cdot 4.49$~GHz and $\omega_{12}=2\pi\cdot4.337$ GHz, which indicates an anharmonicity of $\alpha= -2\pi \cdot 153$ MHz. The spectroscopy is not influenced by AC-Stark shift induced by the drive as we directly probe the transition frequencies. As demonstrated in \cite{Supplementary}, we experimentally verify that the qubit frequency is constant with increasing drive amplitude. This observation shows an additional advantage of our approach and that it performs well over a reasonable range of parameter variations.

\begin{figure}[H]
\centering
\includegraphics[width=0.89\columnwidth]{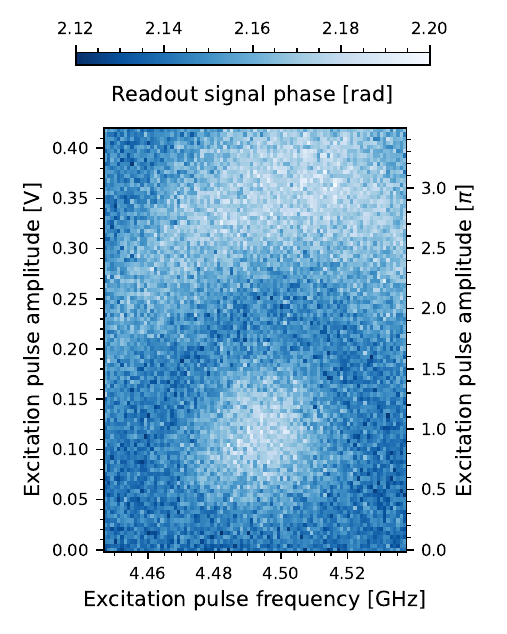}
 \caption{\label{fig:Rabi_chevron} Rabi-chevron response obtained by sweeping the amplitude of the drive pulse with subsequent readout pulse probing the $\omega_{12}$ transition. For technical details see the text. The phase of the probing pulse is shown in the top colorbar.}
 \end{figure}

Next, we have performed time-domain characterization experiments by using the already determined qubit parameters. Transmon anharmonicity limits the minimum pulse duration to $\approx 1$~ns \cite{Koch2007}. To obtain the Rabi-chevron response, we sequentially sent two pulses into the transmission line. The duration of the first drive pulse with a frequency close to the qubit transition frequency $\omega_{01}$ was 24~ns. This duration is long enough compared to the anharmonicity and sufficiently short to capture the dynamics of the qubit. The subsequent readout pulse had a duration of 240~ns at $\omega_{12}$ with a power -125~dBm, which corresponds to 25 photons. These measurements were repeated by changing the amplitude and frequency of the drive pulse. The obtained results are presented in Fig.~\ref{fig:Rabi_chevron}. Other combinations of drive and readout pulses are shown in \cite{Supplementary}. 

\begin{figure}[H]
\centering
\includegraphics[width=0.9\columnwidth]{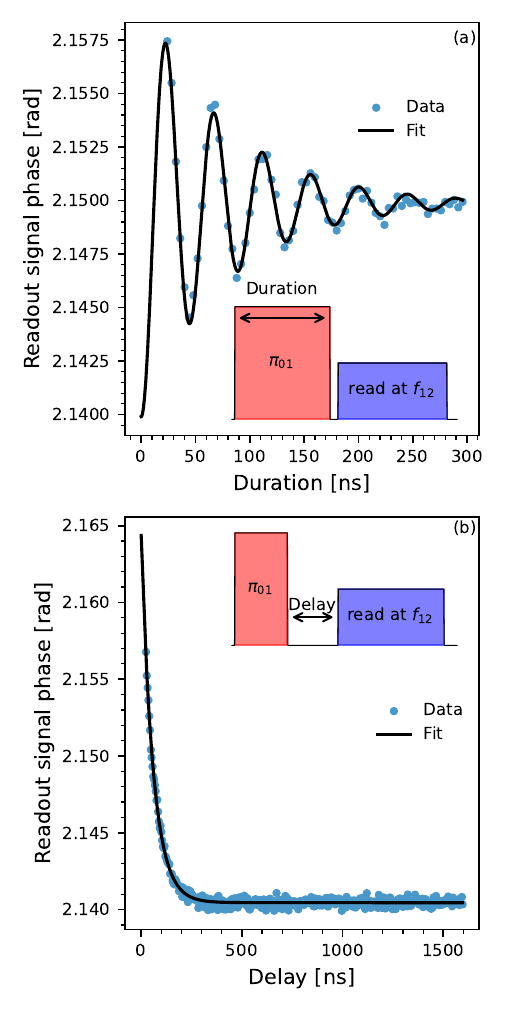}
 \caption{\label{fig:Lifetimes} Qubit evolution. a) Damped Rabi oscillations, obtained by sweeping the drive pulse duration at a Rabi driving rate $\Omega_p=2\pi\cdot 22.47$~MHz; b) Relaxation of the first excited qubit state.} 
 \end{figure}

The scattering of the readout pulse strongly depends on the population of the first excited state, thereby encoding the latter into the amplitude and phase of the detected pulse. In general, by adjusting the readout pulse parameters, one can select where most of the state information will be encoded, either in its phase or magnitude. This is a common approach in cQED. In our case, the phase is more informative and thus shown in Fig.~\ref{fig:Rabi_chevron}, explicitly revealing a standard chevron-like structure that allowed us to define the drive parameters that correspond to the maximum population transfer between the ground and the first excited states. Although we cannot directly infer the actual excited state population value at maximum transfer, we maintain the notation for drive-pulse amplitudes in units of $\pi$ as values for a secondary Y-axis. 
In the second experiment, we swept the duration of the drive pulse at fixed amplitude, corresponding to the $\pi$-rotation. It allows us to estimate the decoherence of the qubit, i.e. the damping rate of the Rabi oscillations $\Gamma_{Rabi}=\frac{\Gamma_{\varphi}+\Gamma_{10}+\Gamma_l}{2}$ and the Rabi drive rate $\Omega_p$. The result is shown in Fig.~\ref{fig:Lifetimes}a, where the fit with decaying oscillations gives us $T_{Rabi}=77.92$~ns and $\Omega_p=2\pi \cdot 22.47$~MHz.

Finally, we directly measured the energy relaxation rate, by varying the delay between the drive $\pi$-pulse and the readout pulse. This rate is defined as $\Gamma_{10}+\Gamma_l$. The result is shown in Fig.~\ref{fig:Lifetimes}b, where the exponential fit gives $T_1=58.86$~ns.

Based on the parameters extracted from these pulse experiments, we estimate the loss rates and compare them with those obtained from the spectroscopic results shown in Fig.~\ref{fig:Qubit_spectra}. The data in Table~\ref{tab:Compare_table} demonstrate good agreement between the standard approach compared to the one proposed herein.
\begin{table}[!t] %
\small  
\caption{\label{tab:Compare_table}Comparison of loss rates obtained using different approaches. All values are normalized to $2\pi$, and values marked as 'n.a.' indicate non-applicability.}
\begin{tabular}{|p{2.3cm}|c|c|c|}
\hline
Parameters & Resonator fit & Qubit fit & Time-Domain \\
\hline
Radiative Loss Rate , $\Gamma_{10}/2\pi$ & 
        $2.26 \; \text{MHz}$ & 
        $2.20 \; \text{MHz}$ & n.a. \\
\hline
Non-radiative Energy Loss Rate, $\Gamma_{l}/2\pi$ & 
        n.a. & 
        $0.31 \; \text{MHz}$ & n.a. \\
\hline
Pure dephasing Rate, $\Gamma_{\varphi}/2\pi$ & 
        n.a. & 
        $1.28 \; \text{MHz}$ &  
        $1.38 \; \text{MHz}$ \\
\hline
Relaxation Rate, $\Gamma_1/2\pi$ & 
        n.a. & 
        $2.50 \; \text{MHz}$ & 
        $2.70 \; \text{MHz}$ \\
\hline
Decoherence Rate, $\gamma_{10}/2\pi$ & 
        $2.62 \; \text{MHz}$ & 
        $2.54 \; \text{MHz}$ & 
        $2.73 \; \text{MHz}$ \\
\hline
Rabi Decay Rate, $\Gamma_{\text{Rabi}}/2\pi$ & 
        n.a. & 
        $1.89 \; \text{MHz}$ & 
        $2.04 \; \text{MHz}$ \\
\hline
Rabi Drive Rate $\Omega_p/2\pi$, used/fit & 
        \parbox[t]{2cm}{\centering $1.06 \; \text{MHz}$} & 
        \parbox[t]{2cm}{\centering $1.06 \; \text{MHz}$ \\ $0.29 \; \text{MHz}$} & 
        \parbox[t]{2cm}{\centering $22.47 \; \text{MHz}$} \\
\hline
\end{tabular}
\end{table}

Additionally, we verify that our assumption of a weak drive, used for the steady-state response, is justified. Specifically, with the drive rate used in the resonator fit \( \Omega_p = 2\pi \times 1.06\,\text{MHz} \), we calculate $
\frac{\Omega_p^2}{(\Gamma_{10} + \Gamma_l) \gamma_{10}}= 0.16 \ll 1. $
This indicates that in the steady state, the transmission at \( \omega_{01} \) should be approximately 0.65, which closely matches the observed dip in Fig.~\ref{fig:Qubit_spectra}a.

We have proposed a simple and versatile method for the characterization of a transmon qubit in an open coplanar waveguide, based on its low anharmonicity. We have demonstrated that this approach enables the extraction of the main parameters of such a system and compared them with the ones estimated by the steady-state approach. Here, we separate the drive and the readout, which provides more flexibility and control over the measurements. 
That allows keeping the first excited state manipulation at the weak-drive limit and use short gate durations, while the readout at $\omega_{12}$ can be optimized. This approach differs from techniques where a low photon number field is used, both for the manipulation and the readout of the qubit's state \cite{Oliver2020, Fedorov2023}. Thus, we believe that our method can be extended and utilized in the research of open quantum systems.

\begin{acknowledgments}
This work was partially supported by the German Federal Ministry of Education and Research under Grant Nos. 13N16152/QSolid, 13N16258/SuperLSI and the European Innovation Council’s Pathfinder Open programme under grant agreement number 101129663/QRC-4-ESP.
\end{acknowledgments}

\section*{Data Availability Statement}
The data that support the findings of this study are available from the corresponding author upon reasonable request.

\bibliography{bibfile}

\twocolumngrid
\clearpage
\onecolumngrid
\clearpage

\renewcommand{\thefigure}{S\arabic{figure}}
\setcounter{figure}{0}

\begin{center}
    \textbf{Supplementary material for "Measuring coherent dynamics of a superconducting qubit in an open waveguide"}  
\end{center}
\section{Sample}
\begin{figure}[h]
    \centering
    \includegraphics[width=0.5\columnwidth]{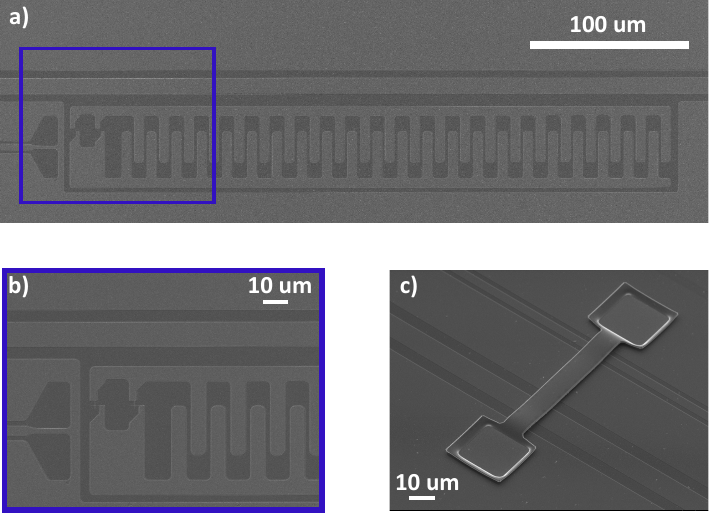}
    \caption{SEM images of the investigated sample: a) the transmon qubit coupled to an open coplanar waveguide, b) zoomed area of the DC-SQUID loop and the DC-bias line, c) the air bridge above the central coplanar waveguide line and the DC-line.}
    \label{fig:sample}
\end{figure}

The investigated sample consists of a transmon qubit capacitively coupled to an open coplanar waveguide. Scanning electron microscope (SEM) images of its basic circuit elements are shown in Fig.\ref{fig:sample}. The open coplanar waveguide has a width and gap size of 10~{\textmu}m and 5~{\textmu}m, respectively. The total length of the transmon shunt capacitor visible in Fig.\ref{fig:sample}a is 383~{\textmu}m. The transmon DC-SQUID loop and the on-chip DC-bias line used to enable tunability are shown in Fig.\ref{fig:sample}b. The DC-SQUID loop has a size of 21 x 14 {\textmu}m$^2$. To suppress unwanted parasitic modes in the waveguide, aluminum air bridges have been implemented. In Fig.\ref{fig:sample}c the air bridge above the central coplanar waveguide line (lower one) and the DC-bias line (upper one) is shown.       
\section{Measurement setup}
The experiments presented in the main text are carried out in a dilution refrigerator with a base temperature of 10 mK. All measurements are conducted in pulse regime using Quantum machines OPX+ system and Octave for up/down conversion. The input microwave signal is attenuated with -40 dB at room temperature and additionally with -70 dB in the dilution refrigerator, distributed over the temperature stages to reduce a thermal population. So, the total attenuation at the sample input is -110 dB. The signal passing through the sample followed by a double junction isolator before it is amplified by 40 dB at the 4K stage with use of low-noise HEMT. At room temperature, the output signal passes through a set of two band-pass filters and is again amplified by a room-temperature amplifier (15 dB) before entering the Octave's RF input. We additionally use the built-in OPX+ amplifier with a 20 dB gain. External flux bias is supplied via a current source and connected through twisted pair wiring to the sample's DC-bias line and additionally filtered by RC-filter at 4K stage.
\section{Readout and drive combinations}
Even though the optimization of the drive and readout pulse parameters is beyond the scope of this paper, we would like to briefly demonstrate the flexibility that arises from the proposed technique.

\begin{figure}[h]
    \centering
    \includegraphics[width=\linewidth]{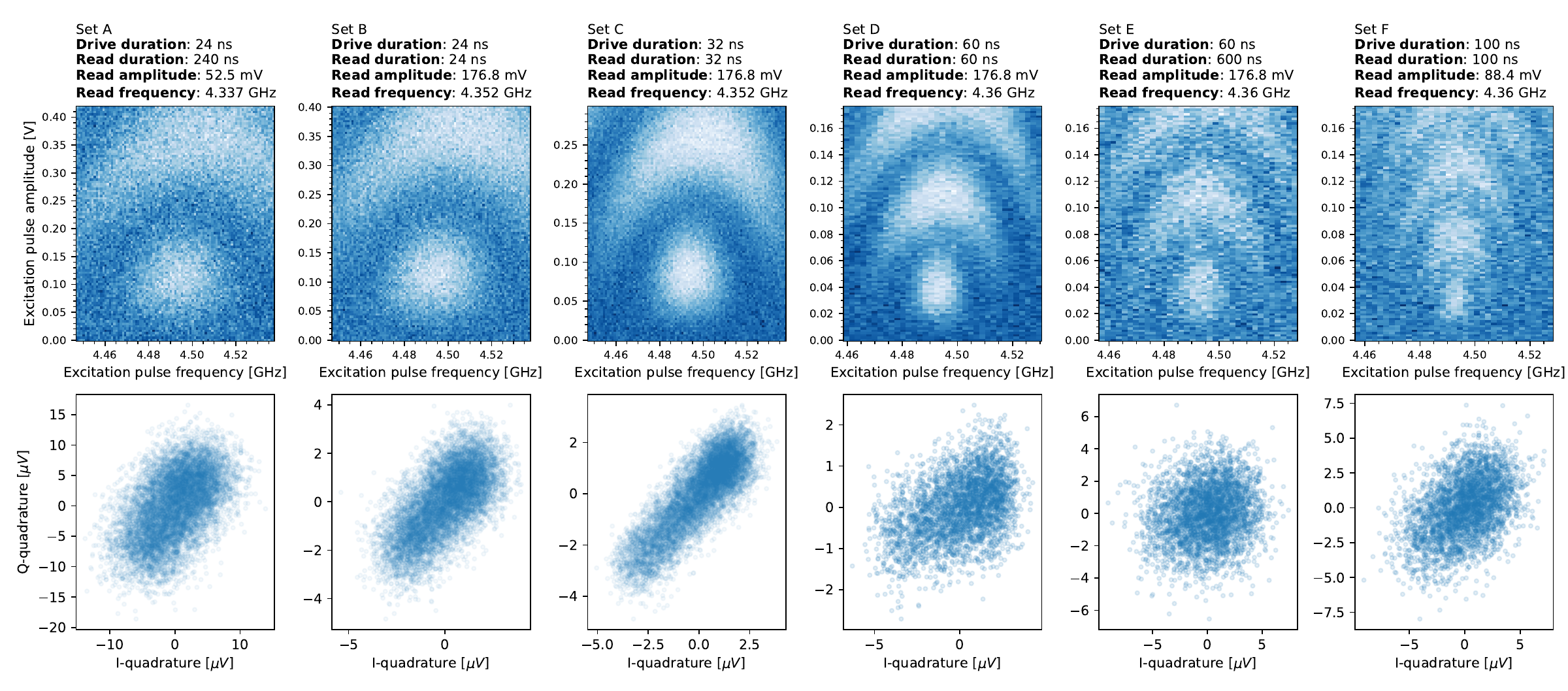}
    \caption{Rabi chevrons measured at different combinations of drive and readout pulse parameters.}
    \label{fig:Different_Rabi_chevrons}
\end{figure}

To gain qualitative insights into the impact of pulse parameters on Rabi-chevron visibility, we performed a series of measurements under varying drive and readout conditions. The different sets for Rabi-chevron experiments are shown in Fig. \ref{fig:Different_Rabi_chevrons}. The parameters of Set A are the same ones used in the experiments presented in the main text.

We observe that shorter drive pulses lead to fewer visible Rabi oscillations in the patterns, as demonstrated by sets A, B, and F. Specifically, Set A (with a 24 ns drive pulse and a 240 ns readout) shows fewer oscillations compared to Set F, where the drive pulse is longer (100 ns). Furthermore, as the drive duration exceeds the coherence times, the contrast diminishes, as seen in sets C, D, and F, where the drive pulses are 32 ns, 60 ns, and 100 ns, respectively. In these cases, the dynamics become smeared out during the drive.

Comparing sets A, B, and C, the measurement results suggest that a shorter readout pulse with a higher amplitude increases the contrast. For example, Set C (with a 24 ns drive and 32 ns readout at 176.8 mV) demonstrates higher contrast compared to Set A (with a 240 ns readout at 52.5 mV). However, it is important to consider the interplay between the readout parameters and the extent to which the readout itself modifies the dynamics of the first excited state, an effect often referred to as back-action.

Additionally, we observe contrast degradation due to relaxation and dephasing during the readout window, as shown by sets D and E, where the readout pulses are 60 ns and 600 ns, respectively. The longer readout durations in these sets result in greater loss of contrast due to relaxation effects. This series of experiments illustrates that, even with non-optimal experimental parameters, the dynamics can still be clearly observed.

Here we also note that the readout frequency was optimized to find the best contrast achievable in the current settings. Moreover, in these measurements most information is contained inside the pulse phase. 
\section{Power dependence}
\begin{figure}[H]
    \centering
    \includegraphics[width=\linewidth]{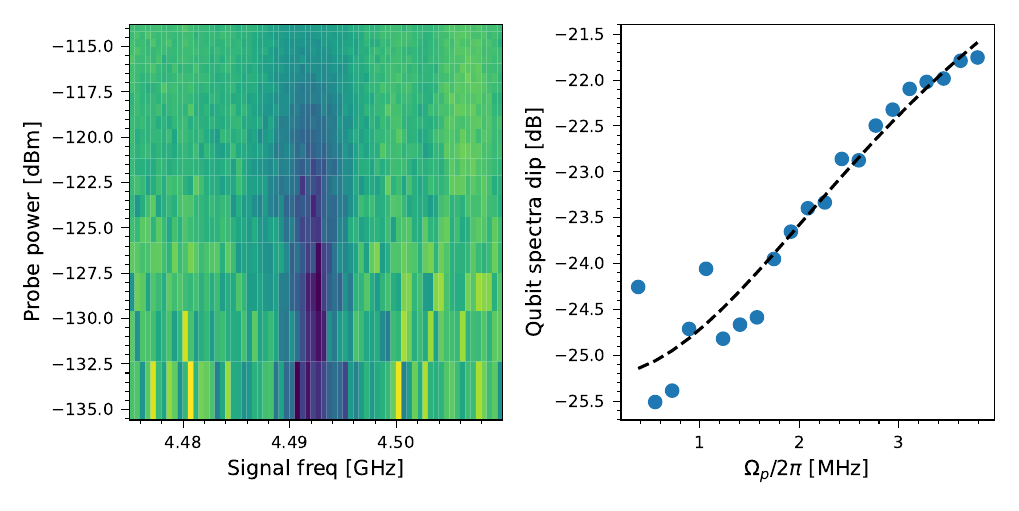}
    \caption{Power dependence of the measured qubit response near the qubit transition frequency $\omega_{01}$.}
    \label{fig:spec_vs_amp}
\end{figure}

The qubit response as a function of the single signal's power is shown in Fig.\ref{fig:spec_vs_amp}, where additionally the dip dependence on Rabi drive rate is shown and fitted with Eq.2 in the main text.

\end{document}